# G-protein coupled receptor subfamily identification using phylogenetic comparison of gene and species trees

Paulo Bandiera-Paiva[1], Jackson C. Lima[1] and Marcelo R.S. Briones[2]

[1]Departmento de Informática em Saúde and [2]Departamento de Microbiologia, Imunologia e Parasitologia, Universidade Federal de São Paulo, 04023-062, São Paulo, Brazil.

To whom correspondence should be addressed
Paulo Bandiera-Paiva
Ed. Leal Prado, Universidade Federal de São Paulo, Rua Botucatu, 862, Térreo.
CEP 04023-062, São Paulo, S.P., Brazil
e-mail:paiva@unifesp.br
Tel: (55)(11) 5576-4347
FAX: (55)(11) 5571-6504

## Abstract

Most approaches to prediction of protein function from primary structure are based on similarity between the query sequence and sequences of known function. This approach, however, disregards the occurrence of gene duplication (paralogy) or convergent evolution of the genes. The analysis of correlated proteins that share a common domain, taking into consideration the evolutionary history of genes under study may provide a more reliable annotation of predicted proteins with unknown function. A computer program that enables real-time comparison of "gene trees" with "species trees" was developed. The Phylogenetic Genome Annotator (PGA) performs a profile based multiple sequence alignment of a set of sequences that share a common domain to generate a phylogenetic gene tree, which is compared to the species phylogeny inferred from aligned ribossomal RNA data. The correlated protein domains are then displayed side-by-side with the phylogeny of the corresponding species. The statistical support of gene clusters (branches) is given by the quartet puzzling method. This analysis readily discriminates paralogs from orthologs, enabling the identification of proteins originated by gene duplications and the prediction of possible functional divergence in groups of similar sequences. The tool was tested in three distinct subfamilies of the G-protein coupled receptor superfamily. In the analysed datasets, the paralogy prediction agreed with the known subfamily grouping, suggesting that subfamily divergence was facilitated by duplication events in the ancestral nodes.

## Introduction

A common approach to function prediction in biology is the use of homology. Biological sequences are compared to other sequences present in databases, which have known function. Based on a statistically significant similarity between the query and a known sequence in the database, homology is presumed and the function of the query sequence is inferred based on the annotated function of the homologous sequence.

The characterization of distinct protein function in sequences sharing a common domain can be improved when considering the evolutionary history of the underlying species compared to the inferred phylogeny of the genes. Differences between these two phylogenies can provide a cue about the occurrence of speciation and duplication events in the genes, suggesting possible new subfamilies for paralogous sequences (Eisen, 1998).

The use of molecular phylogenetics methods may aid in the identification of similar sequences which have the same origin but suffered gene duplication in some moment of their evolutionary history. This can be achieved by comparing the inferred phylogeny of related gene products with the predicted phylogeny of the different species that originated them. Divergences between the evolutionary histories of genes and species may point out possible occurrences of gene duplication events and consequently paralogous sequences in contrast with orthologous genes which were present in an ancestral species and arouse simultaneously to the speciation event. A tool for this phylogenetic comparison which includes a paralogy prediction engine and the visualization of both inferred trees of species and gene evolution is presented.

Classes and subfamilies of the G-protein coupled receptors (GPCR) superfamily were depicted in this study. These receptors consist of a superfamily of cell membrane proteins found in a wide range of organisms. Involved in cellular signaling, this superfamily has been extensively studied by current pharmaceutical research. A very large amount of protein sequence data of these receptors is currently available as a result of several genome projects. Classified according to the receptor's ligand, in classes, subfamilies and subfamily subfamilies, the characterization of a GPCR sequence in one class (or subfamily) may be achieved by sequence comparison to profiles of hidden Markov models or using a basic alignment search tool (BLAST) approach (Altschul et al., 1990), since there is little similarity between classes. Many sequences, although classified in a GPCR class, are considered orphans due to the lack of a conclusively significant similarity with a specific subfamily.

## Methods

A program for the analysis of correlated proteins which share a common domain, considering the inferred phylogeny of the underlying species was developed to identify proteins originated by gene duplication and thus the prediction of possible functional divergence in similar sequences. This program, the Phylogenetic Genome Annotator (PGA), was written in C++ programming language (GNU gcc version 3.2.2) using Qt, a multi platform graphic user interface (GUI) application framework. Application data, such as sequence sets and the PFAM domain profiles are stored in a relational database based on a MySQL database management system. The tool was developed primarily for use in Intel based Linux workstations and can easily be ported to other platforms.

The tool was tested on four distinct classes and subfamilies of G-protein coupled receptors (GPCRs). Three classes of GPCRs were considered, Class A Rhodopsin like, Class B Secretin like and Class C Metabotropic glutamate/pheromone. Class A constitutes the largest class of GPCRs, in order to reduce computing time two subgroups of the amine subfamily were analized, the acetylcholine and serotonin receptors. All receptors from Classes B and C were considered.

The lists of sequences from these classes and subfamilies were retrieved from GPCRDB (Horn et al., 2001). Complete sequences and taxonomic information for each receptor was downloaded from

SWISSPROT (http://www.ebi.ac.uk/swissprot). These datasets were analyzed phylogenetically, comparing inferred phylogenetic trees for the proteins with the underlying species' phylogenies. Using both trees, of genes and species, duplication and speciation events were inferred.

The analysis is based in two coordinate datasets, named sequence sets, of proteins and of ribosomal RNA of the proteins' underlying species. The first dataset is used for inferring the phylogeny of the genes, the former for the inference of the species tree of the source organisms of the proteins from the first dataset.

Sequence sets are data structures that represent groups of correlated sequences, containing taxonomic information, primary structure, and a multiple alignment. This structure is used both for protein sequences, representing gene expression, and for nucleotide sequences of ribosomal RNA, used for the inference of species phylogeny. The gene sequence set is used for the manipulation of protein sequences representing genes of interest, the species sequence set is used for storing the rRNA sequences from the source organisms of the prior set.

The gene sequence set is comprised of a user specified group of related proteins, these sequences may be manually entered, specified by accession id, by a list of accession IDs or automatically fetched from the seed sequences of a specified PFAM protein domain family (Bateman et al., 2000). Proteins specified by accession and those pertaining to the PFAM seed sequences, which don't have complete sequence and taxonomic information, are retrieved by the tool from public databases. All necessary sequence data for inferring both phylogenies, primary structure and taxonomic information, is retrieved either from the NCBI database using XML (http://www.w3c.org/XML) extensible markup language interface or from SWISSPROT using the tool's data communication interface. Lists of accession numbers obtained from GPCRDB for the subfamilies of serotonin and acetylcholine receptors and for class B and C receptors were used for the retrieval of complete sequence information from SWISSPROT.

All user included protein sequences must be related by matching a common protein domain profile or by having significant similarity in their primary structure. Different classes of GPCRs can be easily distinguished since there is no similarity between classes (Bockaert & Pin, 1999), suggesting this superfamily was originated by evolutionary convergence. For the GPCRs groups considered in this study, three protein domain HMM profiles from the PFAM were used, the rhodopsin family (7tm_1, accession PF00001), the secretin family (7tm_3, accession PF00002) and the metabotropic glutamate family (7tm_3, accession PF00003).

Sequences in the gene set are aligned to a hidden Markov model (HMM) profile of a common protein domain. This profile-driven multiple alignment is performed by the hmmalign tool from the HMMER package (Eddy, 2001; Eddy, 1998). Only matching states of the sequences are considered in the alignment. This approach includes only residues which effectively participate in the protein domain being analyzed, and thus represent conserved regions. Using matching states in the sequences of conserved domains renders a more reliable multiple alignment for containing sites which presumably remain constant throughout the evolutionary process.

The underlying species phylogeny is inferred using ribosomal RNA data, which is a universally conserved molecule (Woese, 1970). The species sequence set contains aligned ribosomal sequences relative to the species of the source organisms of the proteins in the gene sequence set. The ribosomal RNA sequences of the species list of the gene set are retrieved from the Ribosomal Database Project (RDP-II) (Cole et al.,2003). The RDP-II is a repository that contains data from small subunit (SSU) ribosomal RNA of eukaryote and prokaryote organisms. Ribosomal RNA sequences in the RDP-II databases are maintained at the repository in an aligned form, these sequences are used for inferring the evolutionary history of the considered organisms. The RDP-II database contains evolutionary representative alignments for prokaryote and eukaryote organisms. For the GPCR datasets, the eukaryote SSU rRNA database was used.

Both sets of aligned sequences, representing the gene set and the corresponding species set, have their phylogeny inferred. Sequences originated from source organisms not present in the ribosomal database are excluded from the gene set in order to provide a strict relation between the two inferred trees in which all protein sequences' source organisms are present in the species tree. The phylogenetic inference for both datasets is performed by the TREE-PUZZLE program package (Schmidt et al., 2002), which uses a quartet-based maximum likelihood phylogenetic analysis, considering gamma distributed rate heterogeneity. The use of gamma distribution for the phylogenetic inference is computationally intensive and may be disabled if processing time is prohibitive. The number of distinct rates and the alpha parameter may also be specified. These phylogenetic inferences generate two unrooted trees, for the studied genes and their underlying species. The phylogenetic inference for the analyzed subfamilies was performed using gamma correction, with the alpha parameter being estimated from the dataset and 8 gamma rate categories.

Since the maximum likelihood phylogenies are obtained by the quartet-puzzling (QP) methods, the QP support, or the frequencies of branches in the space of possible quartets is presented in the maximum likelihood tree. Accordingly, QP values above 70% indicate potentially supported clusters and frequencies above 90% indicate strongly supported clusters. The QP scores can roughly be compared to Bootstrap values although should not be confused with. We recommend the TREE-PUZZLE documentation for a deeper discussion

(Strimmer and von Haeseler, 1996).

Using both datasets a process for inferring paralogy events in the gene tree is performed. The identification of putative gene duplication is based on the method described by Zmasek & Eddy (2001). Since the phylogenetic inference of the gene tree using quartet puzzling with sequences that present large similarity levels tends to produce many polytomies, the algorithm proposed by Zmasek, which presupposes rooted binary trees was modified to extend it's ability to identify paralogies in polytomic trees.

The inference of gene duplication and speciation events uses two polytomic trees, of genes and their underlying species. Given the gene tree **G** and the species tree **S**, every source species from the genes in **G** is present in **S**. The nodes of **S** are sequentially numbered in a preorder transversal, child nodes are assigned numbers always greater than their parents. For each node **g** in **G**, a mapping function **M(g)** is defined as the number of the node **s** which contains all source species from genes harbored in node **g**. Gene duplication events can be inferred on an internal node **g** of **G** by comparing **M(g)** with the mapping function of the child nodes $g_1$ to $g_n$ of **g**, where **n** is the number of child nodes. If **M(g)** is equal to one of the values of **M($g_n$)**, node **g** is considered a duplication event.

## Results

The PGA system is freely distributed according to the General Public License (GPL) in C++ source code or binary format. It may be downloaded at http://compbio.epm.br/pga.

The lists of known receptors from the subfamilies considered in this study retrieved from the GPCRDB, rendered 29 sequences of Acetylcholine receptors, 87 of Serotonin receptors, and 24 of GABA-B receptors. In these datasets, 6 distinct species from Acetylcholine and

Serotonin receptors, and 5 distinct species from GABA-B receptors were present in the Ribosomal Database. The phylogenetic inference of the gene tree considered 21 sequences for the Acetylcholine subfamily, and 35 sequences from the Serotonin subfamily, and 19 for the GABA-B subfamily, since not all proteins in the downloaded datasets had source organisms present in the Ribosomal Database.

Subfamily classification of the sequences within the three subfamilies considered was obtained from the GPCRDB. The phylogenetic inference of the gene tree of the sequence sets presents a consistent grouping of subfamilies according to GPCRDB data. Paralogy prediction also agrees with the known subfamily grouping, suggesting duplication events in the ancestral nodes where subfamily divergence appeared. Paralogy events were also predicted within subfamilies, suggesting the occurrence of gene duplication without functional divergence.

Gene and species trees for the subfamilies considered are presented in the same view by the tool. Inferred gene duplication events are indicated by circles. Subfamilies within the sets were manually labeled. The inferred phylogenies and predicted gene duplication events are shown for the Acetylcholine (fig. 1) and the Serotonin (fig. 2) receptor subfamilies of the class A Rhodopsin like Amine subfamily, and the GABA-B (fig. 3) subfamily from class C, Metabotropic glutamate / pheromone GPCRs.

## Discussion

The use of profiles of protein domain families for the multiple sequence alignments provides a more reliable data set for phylogenetic inference since it implicitly aggregates secondary and tertiary structure information to the alignment. Evolutionary events on genes tend to be more conservative in sites that relate to structures involved in protein functionality. These regions are more appropriate for phylogenetic inferences since the occurrence of many divergences between states in the sequences may lead to incorrect alignments and consequently to the assumption of homoplasy in a non-orthologous site. The seven transmembrane helices domain depicted in the PFAM profile are very conserved within each GPCR family, and thus provide sequence stretches that can be reliably aligned.

The quartet-puzzling method for tree inference is very fast however it tends to generate polytomies in the inferred trees. These polytomies reflect uncertainty in the phylogenetic inference but didn't prevent the detection of paralogy, neither hindered the clustering of proteins by subfamilies in the inferred phylogenies.

Identification of gene duplication events in phylogenies of proteins that share a common domain can provide a means for the refinement of a protein family classification considering it's evolutionary history.

## Acknowledgements

P.B.P. received graduate scholarships from UNIFESP and SPDM(Brazil). This work was supported by grants to M.R.S.B. from FAPESP and CNPq (Brazil), and the International Research Scholars Program of the Howard Hughes Medical Institute (USA).

## Figure Legends

Figure 1. Phylogenetic inference of the Class A rhodopsin like amine subfamily, acetylcholine subfamily, with log likelihood value of -5310.73, presenting a consistent grouping of all 5 types of vertebrate, invertebrate and unknown subfamilies. Paralogy predictions are also very consistent with the subgroups, presenting a discrepancy in vertebrate type 4, where the sequence with accession P17200 of Gallus gallus appears to have diverged differently from the species phylogeny. The low bootstrap value of 66 may indicate an error at the phylogenetic inference. Predicted paralogy at the unknown subgroup may suggest the existence of two distinct groups.

Figure 2. Class A rhodopsin like, amine subfamily, serotonin subfamily inferred phylogeny with a log likelihood value of -8847.59. Also presents an agreement with the different vertebrate and insect subgroups as defined in GPCRDB. The subgroup labeled as 'other' appears clustered to different subgroups. Some paralogy predictions appear within subgroups suggesting gene duplication without functional divergence.

Figure 3. Phylogenetic inference and paralogy prediction for class C, Metabotropic glutamate / pheromone GPCRs, GABA-B subfamily. The inferred tree has a log Likelihood value of -4742.19, presenting clusters for subtype 1, 2 and GABA-B like. Subtype 2 and GABA-B like groups appear to be monophyletic and are separated from subtype 1, with predicted paralogies.

Figure 1.

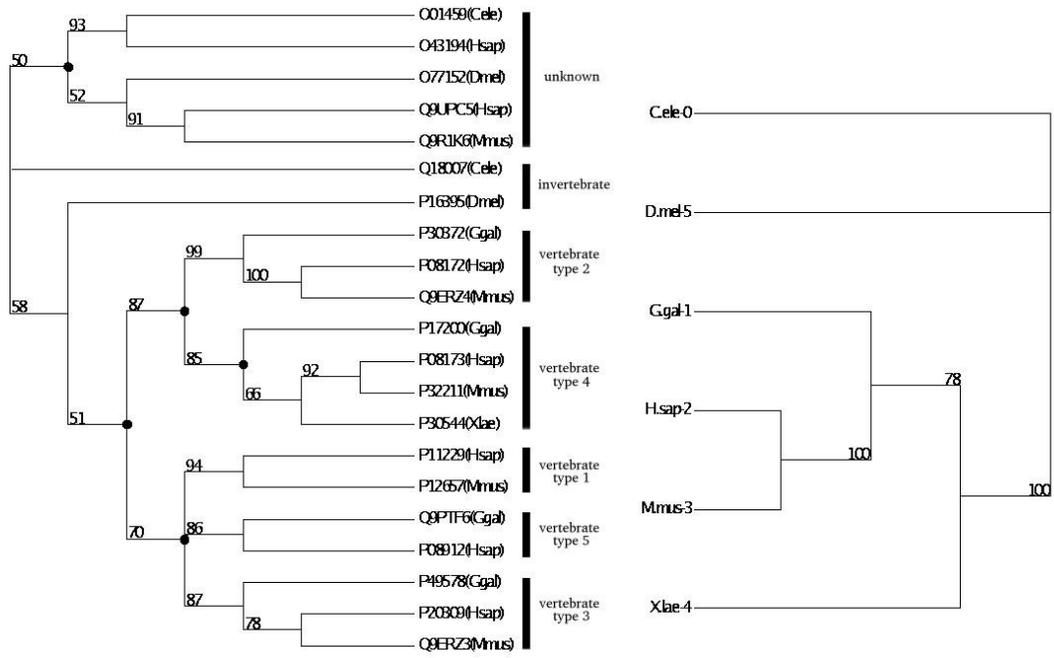

Figure 2.

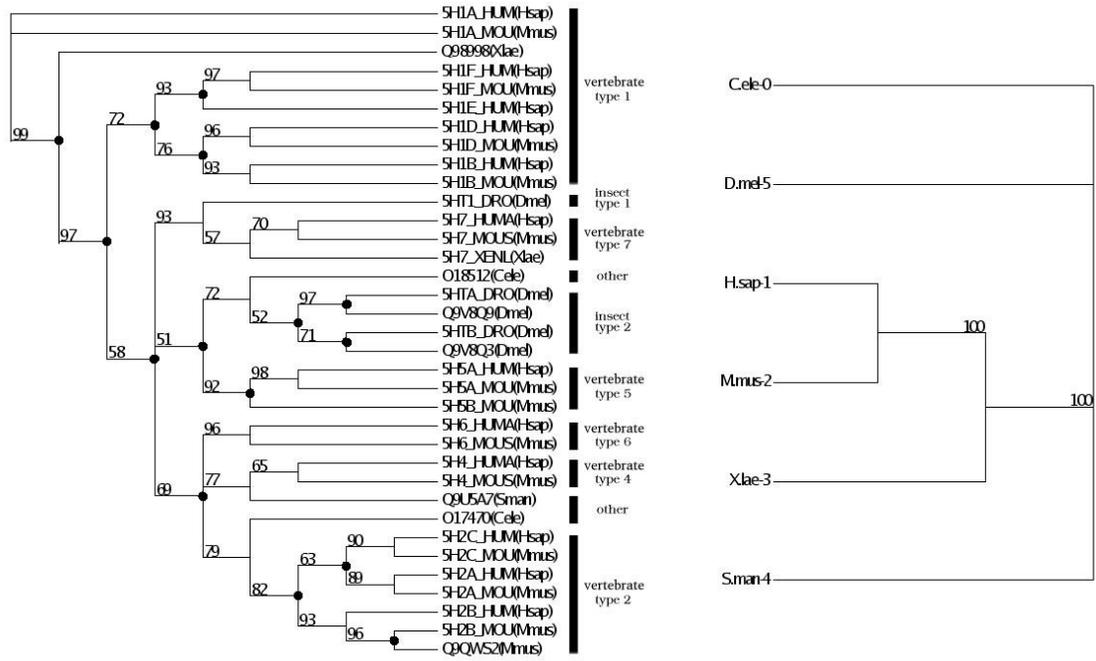

Figure 3.

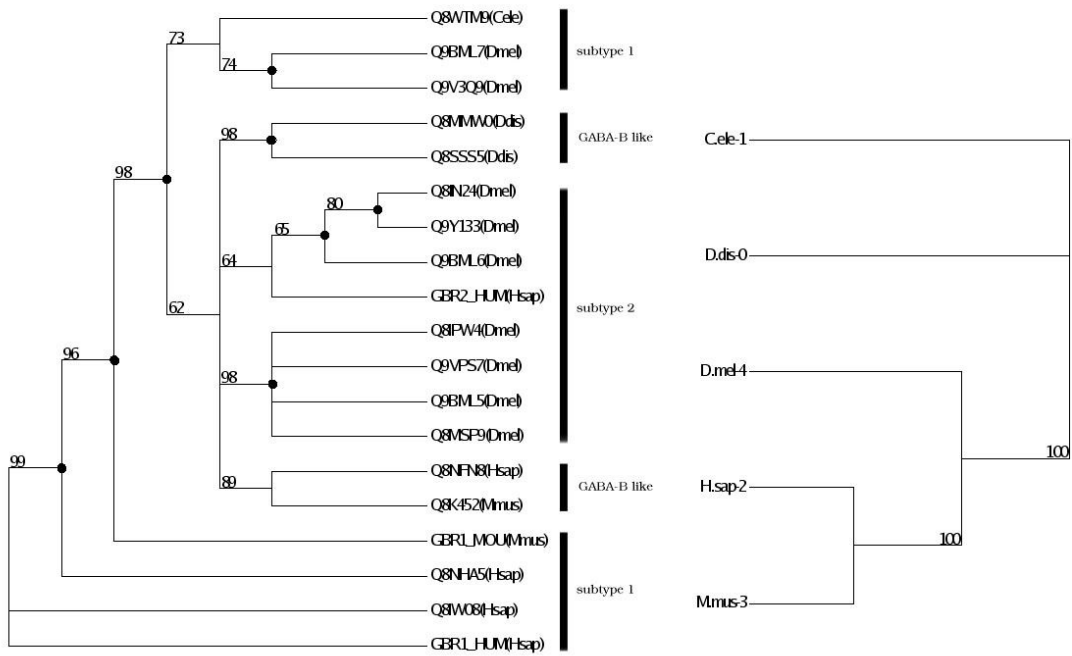